\documentstyle[multicol,prl,aps]{revtex}% for preprint form
\input{epsf.tex}
\date{%Submitted to Phys. Rev. Lett.,  \today
May 25, 2000}
\begin{document}

%\draft
\title{Nanometer Scale Mapping of the Density of States in
 an \\ Inhomogeneous Superconductor}

\author{T. Cren, D. Roditchev, W. Sacks, and J. Klein}

\address{Groupe de Physique des Solides, Universit\'es Paris 7 et Paris
6, \\ Unit\'e Mixte de Recherche C.N.R.S. (UMR 75\ 88), 2 Place
Jussieu, 75251 Paris Cedex 5, France}

\maketitle

\begin{abstract}
Using high speed scanning tunneling spectroscopy, we perform a full
mapping of the quasiparticle density of states (DOS) in single
crystals of Bi${}_{2-x}$Pb${}_x$Sr${}_2$CaCu${}_2$0${}_{8+\delta}$.
The measurements carried out at 5\,K showed a complex spatial pattern
of important variations of the local DOS on the nanometer scale.
Superconducting areas characterized by a well-pronounced
superconducting gap are co-existing with regions of a smooth and
larger gap-like DOS structure. The superconducting regions are found
to have a minimum size of about 3\,nm. The role of Pb-introduced
substitutional disorder in the observed spatial variations of the
local DOS is discussed.

\end{abstract}

\begin{multicols}{2}

The understanding of  how disorder influences the superconducting
state is very important from both fundamental and technological
points of view. Recently, a new insight was gained from numerical
studies of a two dimensional superconductor \cite{Indiens,Huscroft}.
It was predicted that the disorder may destroy the long range phase
coherence, without suppressing the local pairing. In such a
non-superconducting state, the quasiparticle state density (DOS) does
not correspond to the normal metal phase, but to some new one,
characterized by a particular gapped DOS. Based on experimental
observation, the existence of such a phase at low temperature was
suggested \cite{Cren2000}, but it is still an open question how the
superconducting and non-superconducting phases co-exist spatially in
the same two dimensional superconductor.

High temperature superconductors (HTSC) are good candidates for such
a study. They are quasi-two dimensional superconductors, with a large
superconducting gap easy to identify. The quasiparticle DOS, better
known in Bi${}_2$Sr${}_2$CaCu${}_2$0${}_{8+\delta}$ (Bi(2212)), is
characterized by low-lying excitations corresponding to an
anisotropic gap, large quasiparticle peaks, and dips and humps
appearing outside the gap. Such a shape is also a characteristic
footprint of the superconducting state in other (2212) cuprates
\cite{Zasadzinski1993}. Thus, it is a good starting point in the
study of spatial variations of the local DOS, hitherto unknown.

The underlying normal state in HTSC appears complicated, and up to
now the subject of a controversy. Recent tunneling experiments in a
magnetic field (vortex core) \cite{Maggio,Renner1998} or in
disordered thin films \cite{Cren2000} suggested the low temperature
pseudogap (LTPG) to be the signature of such a state. In this sense
also, the study of the spatial evolution of the state density is
important, and a full mapping of the local DOS is needed. Up to now,
however, most of experiments were carried out either as tunneling
spectroscopy measurements in some chosen locations of the sample
surface, or as DOS maps, but performed uniquely at a selected bias.

In this Letter we report a complete DOS mapping of single crystals of
Bi${}_{2-x}$Pb${}_x$Sr${}_2$CaCu${}_2$0${}_{8+\delta}$ with $x=0.5$
(Bi/Pb(2212)), in which Bi-Pb substitution creates an additional
disorder potential throughout the sample. The experiments were
conducted in our home-built low temperature scanning tunneling
microscope (STM) \cite{Mallet1995}. The DOS mapping consists usually
in acquiring a set of current-voltage spectroscopic data $I(V,x,y)$
(or $dI/dV(V,x,y)$ data directly) in every point $(x,y)$ of the
topographic image. At such a point the feed-back loop stabilizes the
STM tip at a given tunneling resistance (fixed $I{}_T$ and $V{}_T$),
and the relative height signal $Z(x,y)$ is used to form a constant
current topographic image. The feed-back loop is then opened ($Z$ is
fixed), the junction voltage is ramped and a set of corresponding
current values is acquired. To enhance the signal-to-noise ratio
either signal averaging or lock-in techniques are used. At the end of
the voltage ramp the junction is biased again at $V{}_T$, the
feedback loop is closed, and the tip height readjustment is made
before moving the tip to the next point of the surface. This data set
may then be presented either as a series of differential conductance
curves at selected locations $dI/dV(V,x_0,y_0)$ or as differential
conductance maps of a region at a given energy $E=eV_0$:
$dI/dV(V_0,x,y)$. In this work, it is crucial that the spectroscopic
data are taken point-by-point simultaneously with the acquisition of
the topographic image, giving an exact spatial correspondence between
them. As a consequence, if the recorded topographic and spectroscopic
images are consistent (i.e. meaningful), then many problems, such as
drift, tip variation, or any other spurious effects complicating the
data analysis, are easily detected and avoided.

The main experimental problem which remains is the acquisition of a
large volume of data in a reasonable lapse of time. For this purpose
we have completely modified our experimental setup. The STM unit is
equipped with a low-noise wide band current pre-amplifier situated
near the tip and working at low temperature. A fully digital
acquisition system, including the feedback loop, is based on a
DSP-integrated acquisition card, allowing fast software piloting of
the experiment. In particular, rapid switching from a topographic to
a spectroscopic mode, with synchronous acquisition, is easily
implemented. Due to these modifications the differential conductance
maps may be obtained in typically 20\,minutes, each $dI/dV(V,x,y)$
set corresponding to 30-65\,MBytes of acquired information.

The samples were cleaved immediately before introducing them to the
loading chamber of the cryostat. The experiments were carried out at
5\,K in a low pressure (10${}^{-3}$\,mbar) of He${}^4$ exchange gas.
Mechanically etched Pt-Ir tips were used.

Contrary to Bi(2212) thin films, in single crystals of Bi/Pb(2212)
the constant current topographic images show a flat sample surface
over large areas without particular growth defects. The apparent
roughness does not exceed 1\,nm (Fig.1a, Fig.2a). However, clear
atomic resolution images were not obtained. A possible reason is that
in Bi(2212) the Bi-O is known to be the surface layer after the
cleaving of the sample. In Bi/Pb(2212) the Bi atoms are substituted
by Pb in these layers, and in arbitrary Bi sites, leading to a weak
disorder at the very surface, masking the atomic pattern.

In many locations the tunneling spectroscopy routinely shows the
superconducting gap in the raw differential conductance spectra, as
Fig.3a shows. This DOS has all characteristic features of the
superconducting state density in $(2212)$ cuprates: large
quasiparticle peaks, dips and humps
\cite{Zasadzinski1993,Mandrus,Renner1995,Mallet1996}, and thus
unambiguously attributed to the Bi/Pb(2212) superconducting phase.
The spectral background is surprisingly flat in the range of
$\pm$300\,mV around the Fermi level, suggesting that within this
region the electronic states from Bi/Pb-O do not give rise to any
gap, additional peak, or any other peculiar signature in the DOS.
Depending on region, the dips/humps appear either symmetrically with
respect to the Fermi energy or only on the occupied states side of
the spectra. The gap width varies somewhat from place to place,
reflecting some variations of carrier concentration. The $d$-wave fit
gives $\Delta=45\pm5$\,mV for the maximum gap value. Consulting the
data on parent Bi(2212) \cite{Zasadzinski1999} and assuming that the
gap/doping dependence in Bi/Pb(2122) is similar, one concludes on an
underdoping of the sample. The clearly pronounced gap, the absence of
a voltage-dependent spectral background along with the small
zero-bias conductance, allows one to rule out surface contamination
as a possible reason for the spectroscopic changes we observed
spatially.

\begin{figure}
%\centering
    \vbox to 8.5cm{
    \epsfysize=8cm
    \epsfbox{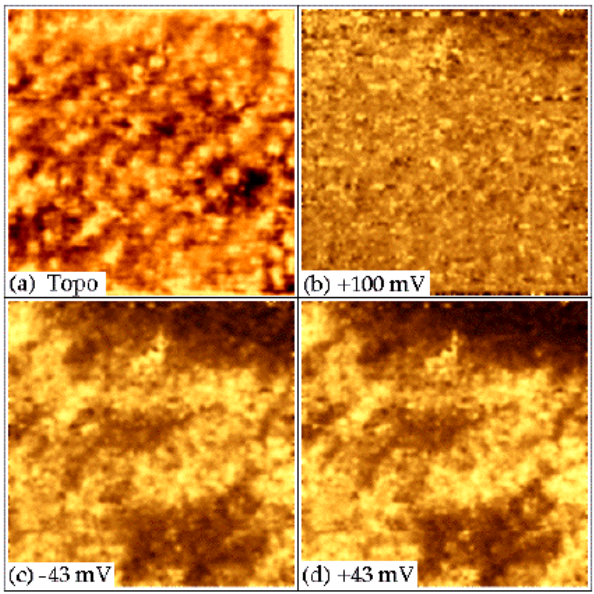}
    }
    %\caption{caption}
\caption{(a) 150 nm x 150 nm constant current topographic image
($V{}_T$= 252 mV, $I{}_T$= 200 pA). The full dark-bright scale
corresponds to 1 nm of height variations.(b) - (d): a set of
corresponding $dI/dV(V_0,x,y)$ maps at different biases $V_0$
(labels). The maps (c) and (d) look identical, representing a strong
spatial correlation in the topology of the quasiparticle peak heights
at occupied (c) and empty (d) state sides of the DOS.}\label{figure1}
\end{figure}

In Fig.1 we show a 150\,nm x 150\,nm topographic image and
corresponding DOS maps at different tunneling voltages. The contrast
in the topographic image corresponds to the apparent tip height as a
function of its lateral position, whereas in the differential
conductance maps the contrast represents a relative magnitude of the
DOS at a given energy. Far beyond the superconducting gap ($|E|\geq
3\Delta$) the DOS is uniformly distributed, as Fig.1b shows. At lower
energies some contrast appears but near the gap edges the maximum
contrast is found, representing the spatial distribution of the
quasiparticle peak heights (Fig.1c for the occupied states side,
Fig.1d for the empty states side of $E_F$). Bright zones in these
images are areas in which the differential conductance has high
quasiparticle peaks. These zones show the DOS having a
superconducting shape, as in Fig.3a,b, and are therefore identified
as 'superconducting' areas. In very dark regions the peaks are
absent, and the state density has a quite different shape (see
Fig.3d). The form of the DOS we found in intermediate regions is
shown in Fig.3c. Such a DOS is identical to the LTPG state density
observed in films of Bi(2212) in our previous work \cite{Cren2000}.
In this Letter we continue to use the term LTPG for the spectrum
having this particular form. Finally, the data in Fig.1 clearly
indicate that the DOS is not homogeneously distributed over the
surface, and superconducting regions are co-existing with
non-superconducting ones. Exploring different locations of the sample
we found similar large (100\,nm and more) superconducting areas.

\begin{figure}
%\centering
    \vbox to 8.5cm{
    \epsfysize=8cm
    \epsfbox{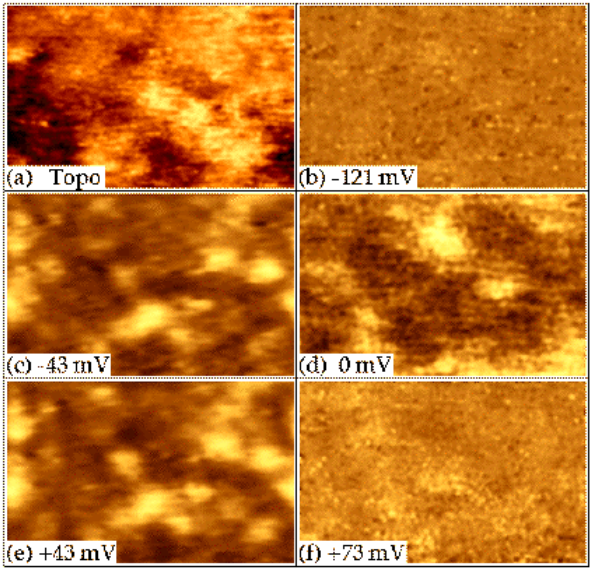}
    }
\caption{(a) 30 nm x 20 nm constant current topographic image
($V{}_T$= 142 mV, $I{}_T$= 100 pA, the contrast scale is the same as
in Fig.1) and a set of corresponding $dI/dV(V_0,x,y)$ maps at
different biases. The maps (c) and (e) at quasiparticle peaks show
the existence of small superconducting regions. Zero-bias conductance
map (d) is roughly inverse of the maps at gap edges (c) and (e). No
contrast is found well outside the gap, either in occupied (b) and
empty (f) states.}\label{figure2}
\end{figure}

In other regions however, the superconducting DOS is not so dominant,
and its variations are observed on a very small scale. In Fig.2 we
present the 30\,nm x 30\,nm topographic image and corresponding DOS
maps of such a typical region. Here the superconducting spectrum is
present only in some locations, forming small islands (Fig.2c). The
position and shape of these islands are precisely the same in the DOS
maps at $E=+\Delta$ and $E=-\Delta$. By analyzing the corresponding
spectra along the points bordering the superconducting islands we
find that they are surrounded by the peak-less LTPG state density (as
in Fig.3c). The transitions from superconducting DOS to LTPG are
spatially continuous, but take place on a very local scale of
1-3\,nm. The analysis of numerous images indicated the existence of a
smallest characteristic size of superconducting islands to be about
3\,nm in diameter. No smaller areas with well-developed quasiparticle
peaks in the DOS were found. Taking into account a relatively high
Fermi velocity $\sim10^7$\,cm/sec reported in \cite{Campuzano1998},
the existence of such small superconducting regions suggests a very
short interaction time of about $\sim3\cdot10^{-14}$\,sec.

The acquisition of a complete spectrum at each point of the image
allows us to correlate unambiguously the DOS maps corresponding to
different energies. In all images, the quasiparticle DOS maps for the
energy fixed at the negative gap edge (as in Fig.1c, Fig.2c)
perfectly correlate to those at the positive gap edge (Fig.1d and
Fig.2d respectively). A rough correlation (a contrast reversal) is
also found between the DOS maps taken at the gap edge and at zero
energy, indicating that the zero-bias conductance is systematically
smaller in superconducting regions as compared to LTPG ones. At the
same time no contrast was found in DOS maps far beyond the gap edge
for both positive and negative biases. In fact, any change in the DOS
at the energy $E=eV_0$ is directly observed in the differential
conductance images $dI/dV(V_0,x,y)$. On the contrary, in topographic
images, the feedback loop controls the tip height using the value of
the tunneling current, which is proportional to the number of states
integrated over the energy window $[E_F,E_F+eV]$. In our experiments
the topographic images were taken at biases far beyond the gap value.
Then, if the spectroscopic changes take place only near the gap edge,
they will appear in such topographic images as a second order effect.
The results in Fig.1 and Fig.2 confirm that the main DOS
modifications take place essentially at the energies near the gap
region, relevant to superconductivity. The conclusion is that the
disorder perturbs the superconducting phase, and is not directly
responsible for any new electronic states appearing in the underlying
'normal' spectral density.

\begin{figure}
%\centering
    \vbox to 7.5cm{
    \epsfysize=7.0cm
    \epsfbox{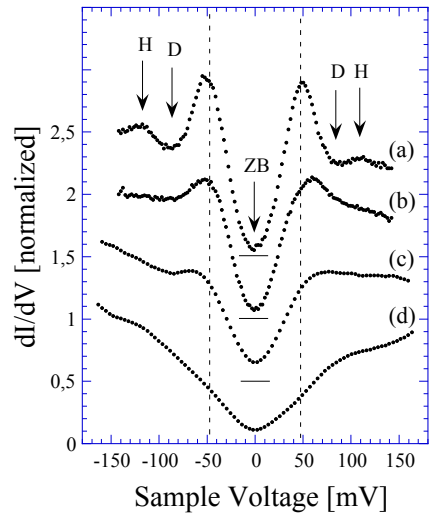}
    }
\caption{State dynamics spatially observed in Fig.1 and Fig.2. (a)
differential conductance spectrum from large superconducting regions
in Fig.1; (b) the spectrum from small islands in Fig.2c,d; (c) the
DOS observed just beyond the superconducting regions; (d) a
structure-less DOS from dark zones of Fig.1c,d and Fig.2c,e (all
spectra are direct numerical derivatives of raw $I(V)$ data). The
spectra are shifted for clarity.}\label{figure3}
\end{figure}

Let us analyse the state dynamics in the transition from
superconducting to non-superconducting phase, as in Fig.3. Starting
from the well-known superconducting DOS (Fig.3a) the very first
effect is the removal of the spectral weight from the quasiparticle
peaks. At the same time the dips/hump features vanish and the zero
bias conductance slightly rises (Fig3c). At a first glance, there is
no conservation of states in such a transformation, since the gap
near the Fermi energy remains, and becomes even a bit larger
\cite{footnote1}. Such a state dynamics in superconducting DOS - LTPG
transition is surprisingly similar to the DOS variations observed in
Bi(2212) single crystals in a magnetic field
\cite{Renner1998,Pan2000}. In these experiments the superconducting
DOS is almost homogeneously distributed over the surface, and
modified only near the vortex cores. There also, the main effect is
the suppression of the coherence peaks on a similar scale (2nm). In
the case of Fig.2 however, the topology of the DOS variations is
opposite to the vortex case: small superconducting clusters are
immersed in a LTPG dominating medium. Further state dynamics is
essentially to enlarge and broaden the LTPG. All sharp DOS structures
disappear completely. Far away from superconducting regions, a wide
and structure-less gap is often observed (Fig.3d). In terms of pair
breaking, the effect would correspond to a strong suppression of the
quasiparticle lifetime. Finally, the state dynamics in the
superconducting DOS - LTPG transition we observe in single crystals
of Bi/Pb(2212) is very similar to that previously reported in
disordered thin films of Bi(2212). Therefore, it is a common effect
in inhomogeneous superconducting cuprates.

In large superconducting regions the differential conductance spectra
show sharp coherence peaks and well-developed dip/hump fine
structure, whereas in the smallest superconducting islands the
quasiparticle peaks are broader and dips/humps are suppressed. The
tentative conclusion is then to associate the dip/hump signatures
with a much larger characteristic scale. Their presence in the
tunneling spectra could be a signature of a long range
superconducting ordering.

The effect of disorder we observed in this work is different from the
result of \cite{Pan1999}, in which Zn-containing single crystals of
Bi(2212) were studied. At low concentration, Zn atoms, situated
directly in Cu-O layers and acting as strong point defects, lead to
the formation of bound states, locally observed. Although the
quasiparticle peaks were found also suppressed near Zn-impurity, the
main effect was a strong conductance peak, appearing near the Fermi
energy. In Bi/Pb(2212) we did not observe such a peak, the main
effect in the DOS being the local suppression of the quasiparticle
peaks. It is not so surprising, since the role of Pb in Bi/Pb(2212)
is very different from that of Zn. In this work the concentration of
Pb atoms was chosen to be $x=0.5$. At such a high density every
fourth Bi atom is substituted by Pb in Bi-O layers, introducing a
quasi-continuous weak disorder potential on the Cu-O layer. With
respect to previously studied thin films of Bi(2212) \cite{Cren2000},
single crystals of Bi/Pb(2212) are much less structurally disordered,
since Pb atoms substitute Bi ones without perturbing significantly
the crystal structure. In this sense, Bi/Pb(2212) single crystals are
much better described by a spatially varying degree of disorder.

To summarize, in this Letter the state dynamics in Bi/Pb(2212) single
crystals is mapped energetically and spatially, giving a direct
answer to recent theoretical considerations \cite{Indiens,Huscroft}.
It shows the differential conductance to be inhomogeneous over the
sample surface on the nanometer scale. The minimum size of a
superconducting island is estimated to be 3\,nm, whereas the spectral
changes take place on a slightly shorter scale (1-3\,nm).
Energetically these variations take place mainly within the range
$[-2\Delta,2\Delta]$ around the Fermi level. The spectral maps taken
within this energy window correlate spatially, whereas they do not
correlate to the topographic information. The effect is the
consequence of the superconducting state - LTPG transition, due to
the continuously varying disorder. The observed topology of the
superconducting regions reveals the existence of two scales: a
nanometer scale, on which the quasiparticle peaks in the DOS are
formed/suppressed, and a larger scale associated with the appearance
of sharp quasiparticle peaks and dip/hump structures. Intuitively,
the first effect would correspond to a very local pairing, the second
one reflecting a larger scale superconducting order.

We are grateful to J. Zasadzinski, J. C. Campuzano and B. Leridon for
stimulating discussions, and F. Breton for his technical assistance.

\end{multicols}


\begin{references}
\bibitem{footnote1}Since in STM the tunneling $I(V)$ curves are normalized with
respect to the tunneling resistance, one can not judge immediately on the absolute
value of the DOS.
\end{references}
\end{document}